\documentclass{article}
\usepackage{myfrascatiphys}
\begin{document}
\title{ 
CHARMED MESON SPECTROSCOPY
}
\author{
Robert K. Kutschke        \\
{\em Fermi National Accelerator Lab,P.O. Box 500, Batavia, IL 60510, USA} \\
}
\maketitle
\baselineskip=11.6pt
\begin{abstract}
In the two years since HQL02, the long sought
$j_{\ell}=1/2$ states have been observed.
In the charmed non-strange sector, these states have the
expected properties but, in the charmed strange sector, the states
have masses below threshold for the otherwise dominant
decay modes, allowing
their observation in suppressed modes.
Improved measurements of the masses and widths of the well
established P-wave charm states have also been published.
\end{abstract}
\baselineskip=14pt
\section{Introduction}

These proceedings will discuss new results in the P-wave sector
of the $c\bar{q}$
systems, where $q$ is one of $u$, $d$ or $s$. The spectroscopy of
these mesons is described
by coupling of the spins of the quark and anti-quark,
$S_{c}$ and $S_{\bar{q}}$, with the orbital angular momentum, $L$, 
between the quark and anti-quark. When $L=1$ this coupling
produces 4 states, with $J^P=\{2^+,1^+,1^+,0^+\}$.
%

Until recently, all of the measured properties of the P-wave sector
were well described by models which exhibit Heavy Quark Symmetry, HQS.
In the limit that the mass of the charmed quark is $>>\Lambda_{QCD}$,
the spin of the charmed quark decouples from the dynamics, 
leaving the total angular momentum of the light quark,
$j_{\ell}=S_{\bar{q}}+L$, as an effective quantum number.  
In this limit, the $c\bar{u}$ and $c\bar{d}$ 
$P$-wave states are grouped into two doublets.  One doublet, with
$j_{\ell}=3/2$, has members with $J^P=\{2^+,1^+\}$; these states decay
to $D^{(*)}\pi$ in a D-wave and have natural widths of order 
20~MeV.  The other doublet, with $j_{\ell}=1/2$  has members with
$J^P=\{1^+,0^+\}$; these states decay to $D^{(*)}\pi$ in an S-wave
and have natural widths of order a few hundred MeV.
In the following the two $J^P=1^+$ states will be denoted as
$D_1$ and $D_1(j_\ell=1/2)$, where the first notation is for
the state with $j_{\ell}=3/2$. To obtain 
the properties of the physical states, the finite mass of the charmed 
quark is  introduced as a small perturbation on the HQS states.

The states with $j_{\ell}=3/2$ are well established and have the
predicted properties.  The states with
$j_{\ell}=1/2$ have only recently been observed and are the
topic of these proceedings.
A more detailed discussion of HQS and a review
of the data up to 10 years ago
can be found in reference~\cite{barteltshukla}.  The experimental
results
reviewed in these proceedings are the first significant new results
since that time so the reference remains relevant.

\begin{sloppypar}
Most models predicted a similar pattern for the $c\bar{s}$ mesons
and the \hbox{$j_{\ell}=3/2$} states do indeed follow the pattern.
There was, however, a model\cite{bardeenhill}
that predicted a rather
different picture for the $j_{\ell}=1/2$ $c\bar{s}$ mesons.
In this model, which combines chiral symmetry with HQS,
the $j_{\ell}=1/2$ $c\bar{s}$ mesons were predicted
to lie below threshold for decay to $D^{(*)}K$. The decay modes
available in this case are:
$D_s^{(*)}\pi^0$, which is isospin violating,
$D_s^{(*)}\pi\pi$, which is is OZI suppressed,
and $D_s^{(*)}\gamma$, which are electromagnetic transitions.
All of these decay modes have small partial widths of,
at most, a few MeV.  
Refer to the transparencies of this talk\cite{mytalk} for a bibliography
of recent theoretical work on the $j_{\ell}=1/2$ $D_s$ states.
\end{sloppypar}

\section{P-wave Charmed Non-Strange Mesons}

There are two new measurements in this sector.
The FOCUS collaboration has presented measurements\cite{focus}
using the traditional method of looking at the
inclusive $D^+\pi^-$ and $D^0\pi^+$ invariant mass spectra.
The second set of new measurements comes from the
BELLE collaboration\cite{belledalitz}, who have pioneered a new
technique, the measurement of excited charm resonances in the
Dalitz plots of the decays $B\to D\pi\pi$ and 
$B\to D^{*}\pi\pi$. 

  In the $D\pi$ mass spectra presented by FOCUS one expects contributions
from five  processes plus combinatoric background.    The five
processes are: $D_2^{*}\to D\pi$, feed-down from 
$D_2^*\to D^*\pi$ , feed-down from $D_1(2420)\to D^*\pi$,
$D_0^*\to D\pi$, and feed-down
from $D_1(j_{\ell}=1/2)\to D^*\pi$.
In the feed-down
processes, the $D^*$'s decay to a $D$ plus unobserved neutrals,
giving a final state of $D\pi$.  Because of the small $Q$ values
in these decay chains, the peaks from the feed-down processes
suffer little kinematic broadening.
In previous inclusive measurements, the first three processes, 
which give rise to narrow peaks, have been well established, 
but the final two processes, which produce broad peaks, 
could not be resolved above the combinatoric background.
%

   Following the earlier experiments,
the FOCUS collaboration first tried to fit their $D\pi$ mass spectra
without including the last two processes.
Their experiment, however, has an order of magnitude higher statistics
than previous experiments and, after trying many models of the
combinatoric background, none was able to produce a good fit to
their data.  Inspection of the residuals of the fits
suggested that the fit would be improved by introducing a 
contribution from a broad resonance.  Such a contribution
was parameterized using S-wave Breit-Wigner\footnote{This work
does not give any information about the $J^P$ of the broad states.
The choice of an S-wave Breit-Wigner was driven by the expectation
that any broad peak would be dominated by the $D_0^*$.}
with a free mass, width and yield.  This contribution is intended
to model the sum of the contributions from an unknown mixture
of $D_0^*$ and feed-down from the $D_1(j_{\ell}=1/2)$.  When this term
was added, the fit produced an acceptable $\chi^2$.  However it was 
never possible to resolve separately contributions from the $D_0^*$
and the $D_1(j_{\ell}=1/2)$.




It has long been anticipated that 
the $e^+e^-$ B-factory experiments would open a new window on 
charm spectroscopy through the analysis
of the Dalitz plots in $B$ decay.  The first hint at the power
of this technique was presented by CLEO\cite{cleo99}, in which they
used a partial reconstruction technique to perform a multi-dimensional
fit to the decay $B^-\to D^{*+}\pi^-\pi^-$.  

  BELLE has presented the first example of this technique
using full reconstruction of the final state.  They presented
a fit to the Dalitz plot of the decay $B^-\to D^+\pi^-\pi^-$
and a 4 dimensional fit to the decay $B^-\to D^{*+}\pi^-\pi^-$.
A key component of their analysis is
that the energy of the $B$ mesons in the $e^+e^-$ center-of-mass
frame is fully determined 
and they require that the energy of their $B$ candidates 
be consistent with this energy.
This requirement removes the feed-down processes
which complicate the FOCUS analysis.  Compared with
the FOCUS data, the BELLE data has an improved
signal to background ratio, at the expense of signal yield.

  The power of multi-dimensional fits is that interference
among the contributing amplitudes gives rise to structures
with distinctive shapes that are readily distinguished from backgrounds.
For example the presence of the $D_0^{*0}$ is established by observing
its interference with the $D_2^*$ in the $D^+\pi^-\pi^-$ Dalitz plot.
Moreover these interference effects are powerful probes of
the $J^P$ of the intermediate states and BELLE establishes
that the $D_0^{*0}$ and $D_1^0(j_{\ell}=1/2)$ states do indeed 
have $J^P=0^+$ and $J^P=1^+$, as expected.
In neither fit does BELLE find a significant contribution from 
a constant amplitude;  that is, the data are fully described by
a sum of resonant contributions, including virtual processes
via the $D^*$ and $B^*$.

  FOCUS and BELLE  also presented new measurements of the parameters
of the $D_2^*$ mesons.  These measurements have errors that 
are comparable to those of the previous world averages.  

\begin{figure}[t]
 \vspace{12.0cm}
\centerline{\includegraphics{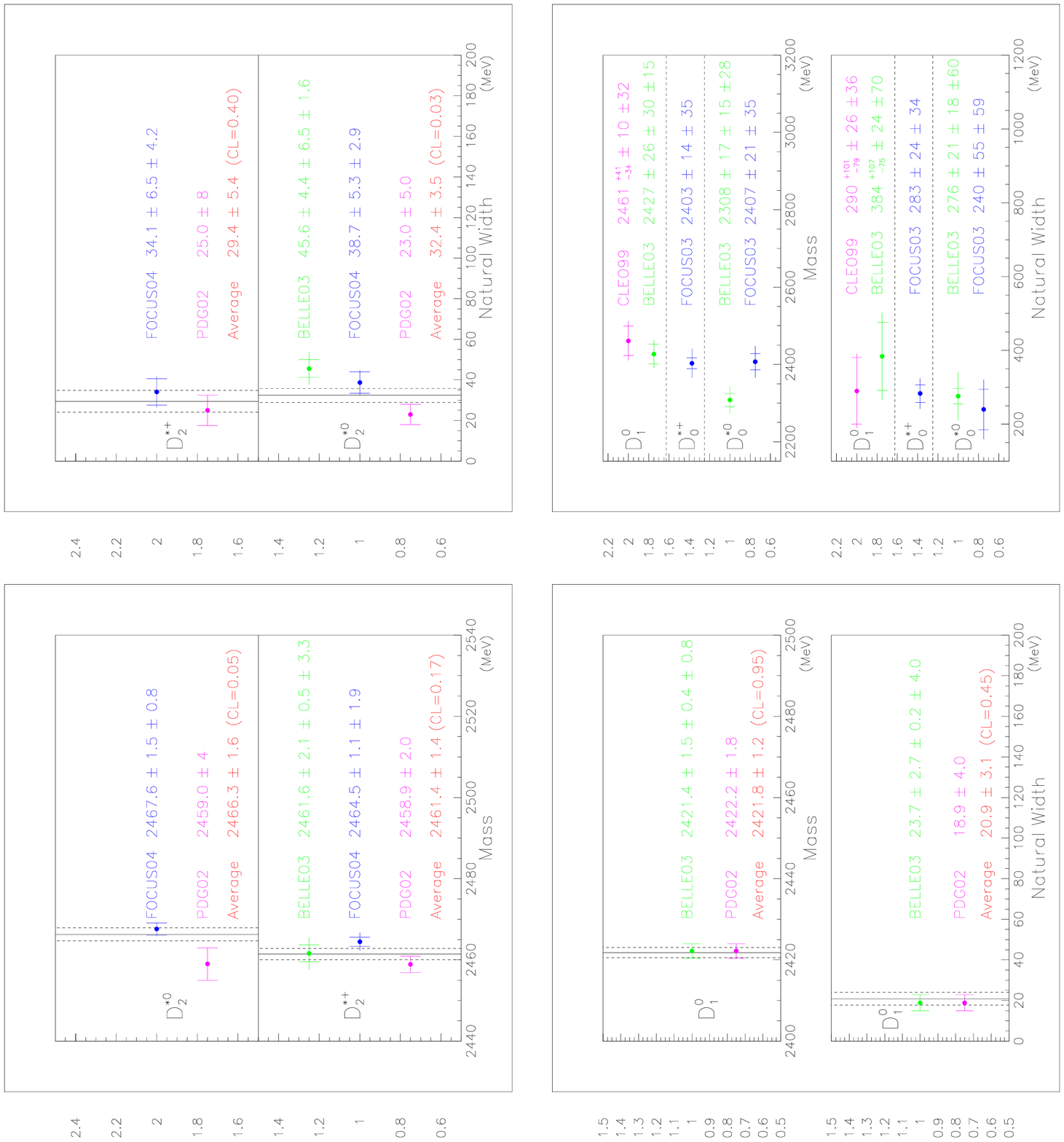}}
 \caption{\it
Masses and widths of charmed non-strange
P-wave mesons.  The first two errors are statistical and systematic.
For BELLE the third error comes from the choice of
contributions to the decay amplitude and for CLEO it
comes from the parameterization of the strong phase.
Parts a) through d) show the results for the well established
$D_2^*$ and $D_1(2420)$ states.  
 Parts e) and f) show the results for the newly observed
broad $D_1$ and $D_0^*$ states.  The averages are taken by the
author; the CL notation gives the confidence level that the
data are self consistent.
As discussed in the text, no average is taken for
the broad states and the BELLE results from the $D_0^*$ should
be be preferred over the FOCUS ones.
    \label{fig:nonstrange} }
\end{figure}

All of the masses and widths discussed above, along with the
PDG 2002 averages and new world averages,
 are shown in figure~\ref{fig:nonstrange}.
Inspection shows that the
new results are consistent with the PDG 2002
values, albeit barely consistent in a few cases.  It does
seem that the new results do prefer broader widths for
both the of the $D_2^*$ charge states.
Perhaps this indicates of a bias toward narrow widths
in early, statistically weaker observations.
Because the FOCUS measurements of the $j_{\ell}=1/2$ states 
are for an unknown mixture of the $D_0^*$ and feed-down from
the $D_1(j_{\ell}=1/2)$,
the author recommends that the best values for the properties
of the $D_0^{*0}$ are the BELLE results alone.

  There are several reasons why the 
B decay results might differ systematically 
from the inclusive measurements.  The line shape of the resonances
is a matrix element squared multiplied by a phase-space factor.
In the inclusive measurements it is difficult to write down the
phase-space factor and it has always been ignored, motivated by the
assumption that it varies only slowly over the region of interest.
In $B$ decays
it is straightforward to write down the phase space factors
and BELLE includes them.
Presumably this is a small effect
for the narrow states but an important effect for the
broad states.  
A second difference is that the inclusive
analyses always assume that the resonances are produced incoherently.
While this is likely, is not certain.

The author looks forward to BaBar entering the field with 
Dalitz analyses of their B decays.  He also looks forward
to both B-factories presenting updated results with much larger
datasets.
In the longer term, both BTeV and LHC-b should
contribute to charm spectroscopy through the Dalitz analysis of
B decays.

\section{P-wave Charmed Strange Mesons}

In the charmed strange sector, the $D_{s2}^*$ and the
narrow $D_{s1}(2536)$ have been well established for more
than a decade.  It was long presumed that the 
the $D_{s0}^*$ and the $D_{s1}(j_{\ell}=1/2)$ would lie above threshold
for decay to $DK$ and $D^*K$.  In such a case this sector would
look much like the non-strange sector, differing only in detail.

  This picture was overthrown when the BaBar collaboration
published the surprising observation of a new, narrow resonance
at a mass of about 2317~MeV which decays to $D_s\pi^0$\cite{babards1}.
Their paper also hinted at a second narrow resonance at a mass near
2456~MeV which decays to $D_s^*\pi^0$.
Shortly afterward the first state was confirmed by
CLEO\cite{cleods}, who also claimed a definite observation of the second state.
Both BaBar and CLEO observed these states in continuum $e^+e^-$ production.
Both states were soon confirmed by BELLE, who observed them both
in continuum $e^+e^-$\cite{bellenonb}
and in $B$ decay\cite{belleb}.
BELLE observed new decay modes of the $D_s(2456)$, to
$D_s\gamma$ and $D_s\pi^+\pi^-$, and 
a new decay mode of one of  the well established states,
$D_{s1}(2536)\to D_s\pi^+\pi^-$.  
BaBar has since confirmed the $D_s(2456)$\cite{babar2}.
Finally, FOCUS has
observed the state at 2317~MeV in $D_s\pi^0$,
which represents the first observation of either state
outside of the $e^+e^-$\cite{vondo}.  
In the following these
states will be refered to the as the $D_{s0}^*((2317)$ and the 
$D_{s1}(2456)$.

  The analysis of the two states is more subtle than is hinted
at by the previous paragraph.  Consider the decay chain,
$D_{s1}(2456)\to D_s^*\pi^0$, $D_s^*\to D_s\gamma$.
If the $\gamma$ is missed and the state is reconstructed
as $D_s\pi^0$, it produces a narrow feed-down peak in the
$D_s\pi^0$ mass spectrum at a mass very close to that of
the $D_{s0}^*(2317)$.  Now consider starting with the decay 
$D_{s0}^*(2317)\to D_s \pi^0$, adding a random photon, requiring
that the $D_s\gamma$ invariant mass fall within the experimental
resolution on the $D_s^*$ mass, and then plotting the $D_s\gamma\pi^0$
invariant mass.  This feed-up process will produce a narrow peak in the 
$D_s^*\pi^0$ invariant mass spectrum at a mass
close to that of the $D_s(2456)$.
A typical mass peak in any of the BaBar, BELLE or CLEO
analysis contains
about 75\% from the signal being looked for and about 25\% from
either the feed-up or feed-down background.  The three experiments
have developed different methods for
unfolding the true signals from these backgrounds and all experiments
get consistent results.

  None of the experiments observe a non-zero natural width for
these states and the best upper limit comes from BELLE~\cite{bellenonb},
$\Gamma(D_{s0}^*(2317))<4.6$~MeV and
$\Gamma(D_{s1}(2456))<5.5$~MeV, both at the 90\% confidence level.

The quantum numbers of these states are already well constrained.
The observation of $D_{s1}(2456)\to D_s\gamma$,
forbids $J=0$ and the BELLE analysis
of angular distributions in the decay $B\to D D_{s1}(2456)$
prefers $J=1$ over $J=2$.
The decay 
$D_{s1}(2456)\to D_s \pi^0$ is not observed, even though phase space
favors it over $D_s^*\pi^0$. This is most easily explained if
the $D_s(2456)$ has $J^P$ from the unnatural sequence,
$0^-,1^+,2^-\dots$.  So the spin parity assignment of $J^P=1^+$
is strongly preferred for the $D_s(2456)$.
 Because the $D_s(2317)$ is observed to decay to 
two pseudo-scalars, and presuming that parity is conserved
in its decay, the $D_s(2317)$ must have $J^P$ from the natural
spin parity sequence, $0^+,1^-,2^+\dots$.  

\section{Summary and Conclusions}

  In the $c\bar{u}$ and $c\bar{d}$ sectors 
 most of the $j_{\ell}=1/2$ states have been observed; only
the $D_1^+(j_{\ell}=1/2)$ remains unobserved.  
These states have the properties predicted by HQS and the
$J^P$ 
quantum numbers are established using the multi-dimensional  
analysis presented by BELLE. 
The results from the old inclusive technique
and the new exclusive technique agree with each other but,
in a few cases, the agreement is only marginal. Perhaps this
is an indication that small effects, which could be ignored
in the past can no longer be ignored in high statistics,
high precision experiments.  The author looks forward to
many years of new results in charm spectroscopy from the
multi-dimensional analysis of B decays.

  In the $c\bar{s}$ sector both $j_{\ell}=1/2$ states have
now been established.  The $D_{s1}(2456)$ has $J^P=1^+$ strongly
favored while the $D_{s0}^*(2317)$ is known to have $J^P$
from the natural sequence, consistent with the expectation of
$0^+$.   Although many people considered the low masses of these states a 
surprise, if you accept the masses then all of the other properties
of these states make sense.  For example the narrow widths arise
because only suppressed decay modes are kinematically allowed.

\section{Acknowledgments}
The author would like to thank the organizers of the conference 
for an exciting program, presented in comfortable and pleasant
surroundings.  He also thanks them for their patience in waiting
for these proceedings.
\section{References}
\end{document}